\documentclass[12pt]{amsart}

\newcommand{\R}{\mathbb R}

\newcommand{\e}{\varepsilon}

\newcommand{\HH}{\mathcal H}

\newcommand{\EE}{\mathcal E}

\begin{document}
\title[Generalizations of wave equations]
{Generalizations of wave equations to multidimensional variational problems}
\author{A. V. Stoyanovsky}
\email{alexander.stoyanovsky@gmail.com}
\address{Independent Moscow University}
\begin{abstract}
This is a survey paper based on previous results of the author. In the paper,
we define and discuss the generalizations
of linear partial differential equations to multidimensional variational problems. We consider two
examples of such equations: first, the generalized Schr\"odinger equation which is a natural candidate
for the mathematical equation of quantum field theory, and second, the quantum Plato problem
which is a natural candidate for a simplest mathematical equation of string theory and, more generally, theory
of $D$-branes. We propose a way to
give a mathematical sense to these equations. 
\end{abstract}
\maketitle

\section*{Introduction}

The purpose of this paper is to introduce the reader to a new area of mathematical physics, called
theory of generalizations of wave equations to multidimensional variational problems. This theory arose
in connection with numerous attempts of scientists to give a rigorous mathematical sense to constructions
of quantum field theory, string theory and, more generally, theory of $D$-branes. Physical intuition behind these
areas of physics already gave rise to many remarkable mathematical results, and so providing a mathematical
background to these disciplines can be very useful for mathematics as well as for physics.

The formal derivation of the generalized wave equations (see \S1 below)
is a direct generalization of formal derivation of
equations of quantum mechanics and wave optics (the Schr\"odinger equation and the wave equation)
from the equations of classical mechanics and geometric optics (the Hamilton--Jacobi equation and the eikonal equation).
The natural examples of generalized wave equations, called the generalized Schr\"odinger equation and the
quantum Plato problem, are given by interesting formulas including determinants.

However, giving a mathematical sense to the generalized wave equations meets problems. It is unclear
what are solutions of these equations. The situation here is similar to the situation at the birth
of theory of partial differential equations and the related famous discussion of scientists in the XVIII century
on what is the mathematical sense of the notion of function. The generalized wave equations are linear
differential equations on infinite dimensional function spaces. But the infinite dimensional
differential operators in these equations, as a rule, {\it cannot be multiplied}. In order to give a mathematical
sense to these equations, one can try to use regularization of the corresponding differential operators. The regularized
functional differential equations have a ``naive'' sense as equations for unknown functionals.
However, in such approach, removing regularization
leads to divergencies. A way to overcome this difficulty essentially worked out by physicists is essentially to subtract
the divergent part of the evolution operator, and consider the remaining finite part; this is
heuristically equivalent to adding an infinite summand to the differential operator.
Thus, in fact, physicists refuse to consider the initial equation, adding to it infinite ``counterterms''.
This construction was usually justified by the argument that in the framework of perturbation theory with respect to
the interaction constant, the differential operator for zero interaction and its corrections in perturbation theory
have no physical sense, and hence may be not considered. However, in our approach to the equations, there is no
perturbation theory at all, and one considers directly the theory with interaction. Therefore this physical
argument looses sense.

The paper consists of two Sections. \S1 is devoted to formal derivation of generalized wave equations.
In \S2 we give an abstract definition of solutions of generalized
wave equations, and discuss the arising problem of constructing appropriate space of
distributions on infinite dimensional function space.

{\bf Acknowledgments.} I thank V. V. Dolotin and V. P. Maslov for illuminating discussions.

\section{Formal derivation of generalized wave equations}

This construction first appeared in the papers [1,2,3] and in the book [4] which can serve as a general introduction
to the ideas of the present paper. The generalized wave equations appeared in these references as purely formal ones,
without definition of their solutions.

\subsection{Generalized Hamilton--Jacobi equation}

Consider the action functional of the form
\begin{equation}
J=\int_D F(x^0,\ldots,x^n,\varphi^1,\ldots,\varphi^m,\varphi^1_{x^0},\ldots,
\varphi^m_{x^n})\,dx^0\ldots dx^n,
\end{equation}
where $x=(x^0,\ldots,x^n)$ is a space-time point, $\varphi^i=\varphi^i(x)$ are real smooth field
functions, $\varphi^i_{x^j}=\partial\varphi^i/\partial x^j$,
and integration goes over an $(n+1)$-dimensional smooth surface $D$
(the graph of the functions $\varphi(x)=(\varphi^i(x))$) with the boundary $\partial D=C$ in the space $\R^{m+n+1}$.
We shall need the well known formula for variation of action on an extremal surface
with moving boundary, which is used, for instance, in derivation of the Noether theorem. Let us recall this
formula. Let
\begin{equation}
x=x(s),\ \ \varphi=\varphi(s),\ \ s=(s^1,\ldots,s^n)
\end{equation}
be a local parameterization of the boundary surface $C$. Assume that for each $n$-dimensional surface $C$
sufficiently close to certain fixed surface, there exists a unique $(n+1)$-dimensional surface $D$
with the boundary $\partial D=C$ which is an extremal of the variational principle (1), i.~e., the graph of a
solution of the Euler--Lagrange equations. Denote by $S=S(C)$ the value of integral (1) over the surface $D$.
Then one has the following formula for variation of the functional $S(C)$:
\begin{equation}
\delta S=\int_C\left(\sum\pi_i(s)\delta\varphi^i(s)-\sum H_j(s)\delta x^j(s)\right)ds,
\end{equation}
where
\begin{equation}
\begin{aligned}{}
\pi_i&=\sum_l F_{\varphi^i_{x^l}}(-1)^l\frac{\partial(x^0,\ldots,\widehat{x^l},\ldots,x^n)}
{\partial(s^1,\ldots,s^n)},\\
H_j&=\sum_{l\ne j}\left(\sum_i F_{\varphi^i_{x^l}}\varphi^i_{x^j}\right)
(-1)^l\frac{\partial(x^0,\ldots,\widehat{x^l},\ldots,x^n)}
{\partial(s^1,\ldots,s^n)}\\
&+\left(\sum_i F_{\varphi^i_{x^j}}\varphi^i_{x^j}-F\right)
(-1)^j\frac{\partial(x^0,\ldots,\widehat{x^j},\ldots,x^n)}
{\partial(s^1,\ldots,s^n)}.
\end{aligned}
\end{equation}
Here $\frac{\partial(x^1,\ldots,x^n)}
{\partial(s^1,\ldots,s^n)}$ is the Jacobian, the hat over a variable means that the variable is omitted, and
$F_{\varphi^i_{x^j}}$ is the partial derivative of the function $F$ with respect to its argument $\varphi^i_{x^j}$.

Let us derive the equations satisfied by the functional
$$
S(C)=S(\varphi^i(s),x^j(s)).
$$
To this end, note that
by formula (4), the quantities
\begin{equation}
\pi_i(s)=\frac{\delta S}{\delta\varphi^i(s)},\ \ H_j(s)=-\frac{\delta S}{\delta x^j(s)}
\end{equation}
depend not only on the functions $\varphi(s)$, $x(s)$, but also on the derivatives $\varphi^i_{x^j}$
characterizing the tangent plane to the surface $D$ at the point $(x(s),\varphi(s))$. These $m(n+1)$ derivatives
are related by $mn$ equations
\begin{equation}
\sum_j\varphi^i_{x^j}x^j_{s^k}=\varphi^i_{s^k},\ \ i=1,\ldots,m,\ \ k=1,\ldots,n.
\end{equation}
Therefore, $m+n+1$ quantities (5) depend on $m(n+1)-mn=m$ free parameters. Hence they are related by $n+1$ equations.
$n$~of these equations are easy to find:
\begin{equation}
\sum_i\pi_i\varphi^i_{s^k}-\sum_j H_jx^j_{s^k}=0,\ \ k=1,\ldots,n.
\end{equation}
These equations express the fact that the value of the functional $S(C)$ does not depend on a parameterization
of the surface $C$.

The remaining  $(n+1)$-th equation depends on the form of the function $F$. Denote it by
\begin{equation}
\HH\left(x^j(s),\varphi^i(s),x^j_{s^k}(s),\varphi^i_{s^k}(s),\pi_i(s),-H_j(s)\right)=0.
\end{equation}
From equalities (5) we obtain the following system of equations on the functional $S(C)$, called
the {\it generalized Hamilton--Jacobi equation} (or the {\it generalized eikonal equation}):
\begin{equation}
\begin{aligned}{}
&\sum_i\frac{\delta S}{\delta\varphi^i(s)}\varphi^i_{s^k}+\sum_j\frac{\delta S}{\delta x^j(s)}
x^j_{s^k}=0,\ \ \ k=1,\ldots,n,\\
&\HH\left(x^j(s),\varphi^i(s),x^j_{s^k}(s),\varphi^i_{s^k}(s),
\frac{\delta S}{\delta\varphi^i(s)},\frac{\delta S}{\delta x^j(s)}\right)=0.
\end{aligned}
\end{equation}
\medskip

{\bf Examples.}
\nopagebreak
\medskip

1) (Scalar field with self-action) [3,4] Let $m=1$ and
\begin{equation}
\begin{aligned}{}
F(x^j,\varphi,\varphi_{x^j})&=\frac12\left(\varphi_{x^0}^2-\sum_{j\ne0}\varphi_{x^j}^2\right)-V(x,\varphi)\\
&=\frac12\varphi_{x^\mu}\varphi_{x_\mu}-V(x,\varphi),
\end{aligned}
\end{equation}
where we have introduced Greek indices $\mu$ instead of $j$, and raising and lowering indices goes using the
Lorentz metric
$$
dx^2=(dx^0)^2-\sum_{j\ne 0}(dx^j)^2.
$$
Then the generalized Hamilton--Jacobi equation has the following form:
\begin{equation}
\begin{aligned}{}
&x^\mu_{s^k}\frac{\delta S}{\delta x^\mu(s)}+
\varphi_{s^k}\frac{\delta S}{\delta\varphi(s)}=0,\ \ k=1,\ldots,n,\\
&D_\mu\frac{\delta S}{\delta x^\mu(s)}+
\frac12\left(\frac{\delta S}{\delta\varphi(s)}\right)^2
+ D^\mu D_\mu\left(\frac12 d\varphi(s)^2+V(x(s),\varphi(s))\right)=0,
\end{aligned}
\end{equation}
where
$$
D^\mu=(-1)^\mu \frac{\partial(x^0,\ldots,\widehat{x^\mu},\ldots,x^n)}
{\partial(s^1,\ldots,s^n)},
$$
and $d\varphi(s)^2$ is the scalar square of the differential $d\varphi(s)$ of the function $\varphi(s)$ on the surface.
\medskip

2) (The Plato problem) [4] Assume that one considers $(n+1)$-dimensional surfaces $D$ with the boundary $C:x=x(s)$
in the Euclidean space $\R^N$
with coordinates $(x^1,\ldots,x^N)$, and the role of integral (1) is played by the area of the surface $D$.
Then the generalized eikonal equation reads
\begin{equation}
\begin{aligned}{}
&\sum_j x^j_{s^k}\frac{\delta S}{\delta x^j(s)}=0,\ \ 1\le k\le n,\\
&\sum_j\left(\frac{\delta S}{\delta x^j(s)}\right)^2=
\sum_{j_1<\ldots<j_n}\left(\frac{\partial(x^{j_1},\ldots,x^{j_n})}
{\partial(s^1,\ldots,s^n)}\right)^2.
\end{aligned}
\end{equation}
In particular, in the case $n=1$, $N=3$, the equation reads
\begin{equation}
\begin{aligned}{}
&x_s\frac{\delta S}{\delta x(s)}+y_s\frac{\delta
S}{\delta y(s)}+z_s\frac{\delta S}{\delta z(s)}=0,\\
&\left(\frac{\delta S}{\delta x(s)}\right)^2+\left(\frac{\delta
S}{\delta y(s)}\right)^2+\left(\frac{\delta S}{\delta z(s)}\right)^2=x_s^2+y_s^2+z_s^2.
\end{aligned}
\end{equation}
\medskip

The generalized Hamilton--Jacobi equation was first systematically studied, for two-dimensional variational
problems, in the book [5]. Regarding the general theory of integration of this equation and, in particular,
regarding the characteristics equations which are the generally covariant generalized
Hamilton canonical equations equivalent to the Euler--Lagrange equations, see [2, 4].

Let us note that in the case of one-dimensional variational problems, integration of the Hamilton--Jacobi equation
is the most powerful method of integration of the canonical Hamilton equations. It would be interesting to check
whether one can integrate the Euler--Lagrange equations for some multidimensional variational problems
(nonlinear partial differential equations, for example, the Einstein or Yang--Mills equations) using integration
of the corresponding generalized Hamilton--Jacobi equations.

\subsection{Generalized wave equations}

Generalized wave equations are obtained from the generalized Hamilton--Jacobi equation (9) by formal substitution
\begin{equation}
\begin{aligned}{}
&\frac{\delta S}{\delta x^j(s)} \to -ih\frac{\delta}{\delta x^j(s)},\\
&\frac{\delta S}{\delta\varphi^i(s)} \to -ih\frac{\delta}{\delta\varphi^i(s)}.
\end{aligned}
\end{equation}
We obtain the following system of linear functional differential equations:
\begin{equation}
\begin{aligned}{}
&\sum_i\varphi^i_{s^k}\frac{\delta\Psi}{\delta\varphi^i(s)}+\sum_j x^j_{s^k}\frac{\delta\Psi}{\delta x^j(s)}
=0,\ \ \ k=1,\ldots,n,\\
&\HH\left(x^j(s),\varphi^i(s),x^j_{s^k}(s),\varphi^i_{s^k}(s),
-ih\frac{\delta}{\delta\varphi^i(s)},-ih\frac{\delta}{\delta x^j(s)}\right)\Psi=0.
\end{aligned}
\end{equation}
The first $n$ of these equations express independence of the ``functional'' $\Psi(C)$ on a parameterization
of the surface $C:x=x(s)$, $\varphi=\varphi(s)$.
\medskip

{\bf Examples.}
\nopagebreak
\medskip

1) In the case of scalar field with self-action (see Subsect. 1.1), we have the following generalized
wave equation which we call the {\it generalized Schr\"odinger equation}:
\begin{equation}
\begin{aligned}{}
&x^\mu_{s^k}\frac{\delta\Psi}{\delta x^\mu(s)}+
\varphi_{s^k}\frac{\delta\Psi}{\delta\varphi(s)}=0,\ \ k=1,\ldots,n,\\
&ihD_\mu\frac{\delta\Psi}{\delta x^\mu(s)}=
-\frac{h^2}2\frac{\delta^2\Psi}{\delta\varphi(s)^2}
+D^\mu D_\mu\left(\frac12 d\varphi(s)^2+V(x(s),\varphi(s))\right)\Psi.
\end{aligned}
\end{equation}
\medskip

2) In the case of the Plato problem, we obtain the following generalized wave equation which we call the
{\it quantum Plato problem}:
\begin{equation}
\begin{aligned}{}
&\sum_j x^j_{s^k}\frac{\delta\Psi}{\delta x^j(s)}=0,\ \ 1\le k\le n,\\
&\sum_j-h^2\frac{\delta^2\Psi}{\delta x^j(s)^2}=
\sum_{j_1<\ldots<j_n}\left(\frac{\partial(x^{j_1},\ldots,x^{j_n})}
{\partial(s^1,\ldots,s^n)}\right)^2\Psi.
\end{aligned}
\end{equation}
In particular, in the case $n=1$, $N=3$, the equation reads
\begin{equation}
\begin{aligned}{}
&x_s\frac{\delta\Psi}{\delta x(s)}+y_s\frac{\delta\Psi}{\delta y(s)}+z_s\frac{\delta\Psi}{\delta z(s)}=0,\\
&-h^2\left(\frac{\delta^2\Psi}{\delta x(s)^2}+\frac{\delta^2\Psi}{\delta y(s)^2}+\frac{\delta^2\Psi}{\delta z(s)^2}\right)=
(x_s^2+y_s^2+z_s^2)\Psi.
\end{aligned}
\end{equation}

\section{On the solutions of generalized wave equations}

In this Section we give a definition of solutions of generalized wave equations.
In this definition, the space-time variables and the field variables are considered in equal rights. This material is new.

Denote by $V$ the Schwartz space
(or any nuclear space [6]) of functions ($\varphi^i(s)$, $x^j(s)$, $\pi_i(s)$, $-H_j(s)$)
(in the case of generalized wave equations (15)), or functions $(x^j(s),-H_j(s))$ (in the case of
quantum Plato problem (17)).
The variables $\pi_i(s)$ are called {\it canonically conjugate} to the variables $\varphi^i(s)$, and the variables
$-H_j(s)$ are called canonically conjugate to $x^j(s)$.
But, for shortness, we shall call these functions simply by $(\varphi(s),\pi(s))$, and omit discrete
indices. Let us introduce the following notation: 
$SV$ is the space of symbols of regular differential operators (the topological symmetric algebra of the space $V$); $SV'$ is the
space of symbols of all polynomial differential operators (polynomial functionals on the space $V$, i.~e., the topological
symmetric algebra of the space $V'$).

Define the {\it Weyl--Moyal} algebra
$SV$ with the Moyal product
\begin{equation}
\begin{aligned}{}
&H^{(1)}*_{Moyal}H^{(2)}(\varphi,\pi)=\\
&\exp\frac{ih}2\int\left(\frac{\delta}{\delta\pi_1(s)}\frac{\delta}{\delta\varphi_2(s)}-
\frac{\delta}{\delta\pi_2(s)}\frac{\delta}{\delta\varphi_1(s)}\right)\ ds\\
&\times H^{(1)}(\varphi_1,\pi_1)H^{(2)}(\varphi_2,\pi_2)|_{\varphi_1=\varphi_2=\varphi,\pi_1=\pi_2=\pi}.
\end{aligned}
\end{equation}

\medskip

{\bf Definition.} Let us call by a {\it general space of distributions} a representation of the Weyl--Moyal
algebra $(SV,*_{Moyal})$
in a topological vector space $\EE$. Let us call by a {\it solution} of a system of differential equations
$(\widehat D_\alpha\Psi=0)$, $D_\alpha\in SV'$,
a distribution $\Psi\in\EE$ such that for any regularization $D_{\alpha,\e}\in SV$, $\e>0$ of symbols
$D_\alpha$, $D_{\alpha,\e}\to D_\alpha$ as $\e\to0$, we have $\widehat D_{\alpha,\e}\Psi\to0$ in the space $\EE$
as $\e\to0$.
\medskip

Numerous examples of general spaces of distributions can be obtained from the unitary representations of the infinite
dimensional Heisenberg group constructed in the book [6] (they are called in [6] by representations of
canonical commutation relations) by passing to the dual space of the space of smooth vectors. Another
example of a general space of distributions is constructed in the paper [7]. See also the book [8]
and references therein. However, up to now no examples (or proofs of existence) of nontrivial solutions
of generalized wave equations are known.

\end{document}